\documentclass[a4paper,11pt]{article}
\usepackage{multicol}
\usepackage[margin=24mm, a4paper]{geometry}
\usepackage{endnotes}

\usepackage{mathptmx}
\usepackage{amsmath}
\usepackage{xfrac}

\let\footnote=\endnote

\begin{document}

\begin{multicols}{2}[\noindent\textbf{\Large Frequentist vs Bayesian statistics --- a non-statisticians view}\\ \\ Maarten H. P. Ambaum \\
\textsl{Department of Meteorology, University of Reading, UK}\\
July 2012]

\bigskip\noindent People who by training end up dealing with probabilities (``statisticians'') roughly fall into one of two camps.  One is either a \textsl{frequentist} or a \textsl{Bayesian}.  To a scientist, who needs to use probabilities to make sense of the real world, this division seems sometimes baffling.  I will argue that science mostly deals with Bayesian questions.  Yet the dominance of frequentist ideas in statistics points many scientists in the wrong statistical direction. 

Frequentists define probability as the long-run frequency of a certain measurement or observation.  The frequentist says that there is a single truth and our measurement samples noisy instances of this truth.  The more data we collect, the better we can pinpoint the truth.  The archetypal example is the repeated tossing of a coin to see what the long-run frequency is of heads or tails; this long-run frequency then asymptotes to the truth.

Bayesians define probability as the plausibility of a hypothesis given incomplete knowledge.  This is in fact the original definition of probability.\footnote{See, for example, J. M. Keynes, \textsl{A treatise on probability.} Macmillan, London (1921); E. T. Jaynes, \textsl{Probability theory. The logic of science.} Cambridge University Press, Cambridge (2003).}  To a Bayesian there is no Platonic truth out there which we want to access through data collection (or perhaps we should say there may be a Platonic truth, but it will always remain outside our experience).  For a Bayesian, there is just data which we can use as evidence for particular hypotheses.  A Bayesian coin-tosser just observes a series of coin tosses and then uses this information to make deductions about, for example, how likely it is that the coin is fair.

The coin-tossing example is in fact quite subtle.  It may be argued that the frequentist view of the experiment appears fine in some Platonic sense but is flawed in practice.  We need to ask ourselves whether there is a ``true'' fraction to which the frequency in the coin-tossing experiment for a fair coin should asymptote.  We have a specific, but idealised view of what the repeated experiment should entail.  We would probably argue that all coin-tosses happen under the same circumstances, but in such a way that the outcome is essentially maximally random.  There is a contradiction here: the coin-tosses are the same insofar that each experiment must have the same validity and samples the same set of possible outcomes in the same way.  But equally, the coin-tosses have to be different in some sense, otherwise the outcomes would be the same.

What is an acceptable coin-toss in reality?  We can agree it is not just dropping a coin from two inches high.  We would need to give it a good spin and drop the coin from a good height.  But we cannot move away from the fact that the experiment is ultimately deterministic.  So, at least in principle, the outcome of the coin-toss is fully determined by the initial condition.

So we see that the long-run frequency for a real coin-tossing experiment is as much a function of the way the experiment is performed as it is a property of the coin.  If we define the initial conditions precisely then the outcome of the coin-toss can be calculated precisely (at least, in principle).  So apparently we need to allow for some variability, some uncertainty in the experimental set-up.

This is the frequentist trap: if we get an asymptotic frequency of \sfrac{1}{2} for heads to come up, is this evidence that we have done a truly randomized (whatever that may mean) experiment to discover that the coin was fair, or is it evidence that we have done a biased experiment with an unfair coin?  We cannot distinguish these by a frequentist experiment, because to distinguish between the two cases we need to know precisely how the experiment was done.  But this in a way precludes the frequentist notion of a repeated experiment.

Instead, the real-world coin-toss experiment lives in the Bayesian domain: the experiment is done under limited knowledge of the initial conditions and precise set-up.  Given that limited knowledge, how do we use the experimental outcomes to evaluate a statement such as ``the coin is fair''?  This is a Bayesian question.  Bayesian statistics is the statistics of the real world, not of its Platonic ideal.

Let us look at an analysis of a particular experiment where 20 coin-tosses were performed   and we got 12 heads and 8 tails (we shall ignore the order in which the heads and tails came up).  What next?  We shall first take a frequentist view on this and then a Bayesian view.

The frequentist sees the particular outcome as an instance of an infinity of possible outcomes, given a particular truth.  Suppose we start by thinking that the coin is fair (that is, in our idealized  experiment we expect the long-run frequency to asymptote to \sfrac{1}{2}).  The frequentist would then calculate probabilities of the particular outcome, given the truth.  For example, what is the probability of 12 heads and 8 tails if the coin is fair?  It turns out to be a quite respectable 12\%.  The chance of getting 11, 10, or 9 heads (that is, closer to the ``truth'') turns out to be 50\%.  So we have done worse than being within one coin-toss away from the truth, but the chance of doing so is 50\% as well.  This becomes more pertinent if we find values that are further away from the expected truth.  For example, suppose we get 13 heads and 7 tails.  Then there is about a 74\% chance of getting closer to the truth than this if the coin was fair.

Should we now start to suspect the coin?  This question is impossible to answer without moving to the Bayesian domain. The chance of getting 13 heads and 7 tails with a fair coin is about 7.4\%.  But suppose the coin (or the experiment ---there is no way to distinguish here) was biased and the true long-run frequency for heads was \sfrac{13}{20} instead of \sfrac{1}{2}.  In this case the chance of getting 13 heads and 7 tails is about 18.4\%, a factor of 2.5 higher compared to the fair coin.  But even though our experimental evidence is more consistent with a biased coin, we probably would still not think that our coin is likely to be biased; we would probably think that the coin was fair and that this experiment just came out this way by chance.  But to come to such a conclusion we actually have to include some prior idea that it is more likely that the coin is fair. In other words, to assess the question of the fairness of the coin, in light of the frequentist evidence that we got 13 heads and 7 tails, we need to have access to the non-frequentist prior information about how likely we think it is that the coin was fair to start with.

A Bayesian sees one particular experiment and uses this to test some hypothesis.  So we can ask the question ``what is the probability that the coin is fair, given that I got 13 heads and 7 tails?''  Using Bayes' theorem this is written as a combination of conditional probabilities:
\[
p(\text{fair}|\text{13h7t}) = \frac{p(\text{13h7t}|\text{fair})\,p(\text{fair})}{p(\text{13h7t})}.
\]
We see that the familiar probability $p(\text{13h7t}|\text{fair})$ of getting 13 heads and 7 tails assuming the coin is fair (which is 7.4\% ---see above) needs to be combined with prior probabilities on whether we think the coin is fair to start with, $p(\text{fair})$, and the prior probability of getting the outcome of 13 heads and 7 tails, $p(\text{13h7t})$, irrespective of any knowledge of the fairness of the coin.  This last probability is hard to assess, but it could in principle be calculated by summing over all possible complementary truths $i$ (with ``fair'' being just one of those possibilities):
\[
p(\text{13h7t}) = \sum_i p(\text{13h7t}|i)\, p(i).
\]
We can view it as the normalization of $p(\text{fair}|\text{13h7t})$ over all possible truths.  Either way, it is not necessarily easy to get good values for these prior probabilities.  In this sense, the Bayesian viewpoint appears to add difficulty for not much gain.  However, it does make explicit what prior knowledge is required before an assessment of the probability of the truth of a hypothesis can be calculated.

Although the example above is somewhat starkly observed, it should make clear that the frequentist viewpoint cannot tell much useful about the hypothesis of whether a coin is fair.  It should be emphasized that a frequentist would, or should not claim to give such evidence.  The frequentist gives a probability of an event given a truth (the $p(\text{13h7t}|\text{fair})$, above) and tries to use this information for any statements. 

Such frequentist statements are the basis for significance testing.  These statements cannot say much useful about the validity of the underlying hypothesis, even less, anything quantitative.  Indeed significance tests formally do not aim to do so, although the terminology of significance tests is notoriously misleading and obfuscating ---for example, a ``significant'' difference between two measurements means something very specific, but, contrary to what the phrase implies (and, unfortunately, contrary to how significance tests are mostly used in practice), it does not mean that the difference has a high probability of being significant.\footnote{There is a long history of people pointing out the terrible misuse and misinterpretation of significance tests in all areas of science. In atmospheric science: N. Nicholls, 2001: The insignificance of significance testing. \textsl{Bull.
Amer. Meteor. Soc.}, \textbf{82}, 981--986; M. H. P. Ambaum, 2010: Significance tests in climate science. \textsl{J. Climate}, \textbf{23}, 5927--5932. In psychology: J. Cohen, 1994: The Earth is round ($p<0.05$). \textsl{Amer. Psychol.}, \textbf{49}, 997--1003; J. E. Hunter, 1997: Needed: A ban on the significance test. \textsl{Psychol. Sci.}, \textbf{8}, 3--7. In economics: S. T. Ziliak and D. N. McCloskey, \textsl{The Cult of Statistical
Significance.} University of Michigan Press (2008). The list goes on.}

\textsl{A frequentist can calculate probabilities precisely, but often not the probabilities we want. A Bayesian can calculate the probabilities we want, but often cannot do it precisely.}

The above characterisation is rather crude, and can be criticised in all kinds of ways, but it does give the flavour of the key practical difference between the viewpoints. There are clearly more precise differences, most importantly in what ``probability'' actually means. But to argue about the different merits of the two viewpoints can easily descent into sophistry. (Clearly, this essay itself is in some danger to be an exercise in sophistry.)

In practice we end up mixing the two viewpoints and we should choose the most appropriate one. But given the historical dominance of frequentism in teaching statistics, it is worth advocating that Bayesian statistics is in many ways a more  fundamental, and more useful view of statistics.

Firstly, we need to dispel the myth that a Bayesian probability, the plausibility of a hypothesis given incomplete knowledge, is in some sense a more vague concept than a frequentist probability, which is based on counting possible outcomes. It has been rigorously shown, most notably by Richard Cox\footnote{See, for example, R. T. Cox: \textsl{The algebra of probable inference.} The Johns Hopkins Press, Baltimore (1961).}, that there is nothing inexact about Bayesian probabilities: they must satisfy precisely the same algebraic rules as frequentist probabilities. In other words, Bayesian probability has as powerful an axiomatic framework as frequentist probability, and many would argue it has a more powerful framework.

Another myth to dispel is that Bayesian statistics is too advanced for basic statistics teaching. Your mileage may vary, but I can honestly state that when I first came across probabilities in mathematics class, I needed to change my essentially Bayesian idea of probability into a frequentist one. After all, in daily usage, statements of probability are overwhelmingly Bayesian, and not frequentist: ``How likely is it I will get wet on my way to school?'' ---an example of an everyday, perfectly sensible question about probabilities which is nearly impossible to interpret sensibly in a frequentist framework.

This begs the question of how we approach traditional frequentist teaching examples within a Bayesian framework. For example, we have a bag with 10 red and 20 blue marbles and draw one marble out of the bag. What is the probability of drawing a red marble? In traditional teaching we have approached this kind of question through a frequentist framework: we imagine we can repeat this experiment as often as we like; it is then clear that we can draw any of the marbles, irrespective of colour, and so in \sfrac{1}{3} of the cases we would draw a red marble. So the long-run frequency of drawing a red marble would be \sfrac{1}{3} and this then is our desired probability.

The Bayesian view invites much more interesting questions about related set-ups (``I just drew a red marble; what is the probability that the majority of marbles is red?'') but in the present boring set-up the Bayesian viewpoint is essentially the same as the frequentist one but more directly related to the real-world experiment: I am going to draw a single marble; in the Bayesian viewpoint it is then a perfectly sensible question to ask what the probability is that in that single particular draw I get a red marble. The answer is clearly the same: one in three marbles is red and so this probability is \sfrac{1}{3}. The Bayesian interpretation of probability of a single event is completely natural: given my knowledge of the experimental set-up (there are 10 red marbles and 20 blue marbles in this bag), what is the probability that we get a particular outcome. The frequentist apparently needs to add the thought experiment of repeating the experiment an infinite number of times. So I ask you: which viewpoint more easily applies to this textbook example?

In all honesty, I cannot think of a single example in science where a frequentist viewpoint is more natural than a Bayesian viewpoint, although I am keen to hear of such examples, as they perhaps do exist. Think of the repeated measurement of some variable; to make things specific, think of measuring the period of a pendulum with a stopwatch. The frequentist would argue that repeated measurements of this period would represent a sample from some true distribution of possible outcomes of this measurement. Our repeated measurements can then be used to make statements about this underlying distribution. A typical statement would be that the best estimate of the mean of this underlying distribution is the average of the measured periods.\footnote{I do open a can of worms here: if I recorded the frequency instead and would take the average frequency to be the best estimate of the frequency of the pendulum, this would correspond to a different outcome than if I took the average of the periods, and then took the inverse. To solve this terrible conundrum, we need to know which of these two variables are more closely described by a simple symmetric distribution, something that is highly dependent on the precise experimental set-up, and also very hard to assess with a limited set of measurements. Thankfully, if the spread is small enough this makes not much difference.

In more complicated fields of science, this is not an easy issue to address and is often simply ignored. For example, many people happily talk about the average daily rainfall without asking what this average actually represents. Daily rainfall has a highly skewed distribution, and the average of a set of samples is non-trivially related to the shape of this distribution.  Average daily rainfall \textsl{does not} measure the most likely rainfall value you might get on any one day ---you mostly get rainfall days with very little rain and every so often days with (much) more rain. In fact, even if correctly interpreted as the scale variable in the appropriate rainfall distribution, a simple average daily rainfall is not the right estimate of this scale. Let us keep the can of worms closed for now.}

The Bayesian approach to the same experiment to my mind is more natural. We have this set of measurements of the period. With this information we can assess the probabilities of all kinds of statements. For example, what is now the probability that if I make another measurement the period of this pendulum is between such and such a value? Bayes' equation again gives the practical way of approaching this (and at the same time reminds us that there are all kinds of prior assumptions going into this experiment), but more fundamentally it does not require us to assume the presence of some ``true'' underlying distribution from which we are sampling. The pendulum will be subject to all kinds of unknown environmental factors and we know little about the underlying distribution (although the central limit theorem may help us in some cases). Instead, we measure some values with some spread, and we can use Bayes' theorem for example to decide on the most likely distribution of periods given what we know about the experiment. In fact, if we only know these measured values, and nothing else, then the most likely distribution is a Gaussian with the measured mean and spread. This is a result from making certain assumptions on prior distributions, for example following a maximum entropy argument; whatever one's views about these techniques, they do amplify the fact that most interesting, real-world probabilistic statements rely on prior assumptions. In a frequentist approach these assumptions are mostly implicit and sometimes hidden; in Bayesian statistics these prior assumptions are made explicit.  The fact that we do not resort to some Platonic ideal outcome from fictional repeated experiments, but deal with the world as it is presented to us, warts and all, to me gives Bayesian statistics the edge.

Up to this point I have deliberately discussed examples which would at first sight perhaps be most naturally viewed as frequentist problems. However it was found that a Bayesian point of view is at least as natural a way of considering these examples. Moreover, real science presents us often with questions that are of a probabilistic nature, but definitely not of a frequentist nature. ``How likely is it that climate change is man-made?'' There is no sensible way in which this can be viewed as a frequentist problem. There are also questions that at first sight do not appear to be of a probabilistic nature but in practice they still are. ``Does medicine A work better than medicine B?'' This question seems to be of a  deterministic rather than a probabilistic character. However, given that a real-world  medical trial would normally present us with evidence that is not 100\% conclusive, any statements made about the relative efficacy of the two medicines will be a probabilistic statement. In fact, most of science appears to be like that.\footnote{This is clearly only a very partial picture of science. Think of Newton's laws of gravity. In some sense we are certain they are incorrect: Einsteins theory of gravity is a better model of gravity. Such a situation is not obviously described in probabilistic terms, frequentist or Bayesian, except in a trivial way. In physics we are usually quite certain that all our models are wrong at some level of scrutiny but are right at another level. The standard model of particle physics is very likely wrong, or at least incomplete, yet it has up to now survived all experimental scrutiny. I think physics is in this sense special: there appear to be grand narratives, theories that work in one situation and not in another. However, statements in physics about individual experiments become probabilistic again; think of the recent discovery (or not) of the Higgs boson.} Science is not like mathematics; there are no absolute truths or falsehoods. We just gather evidence that supports certain models of reality.

To end with a Bayesian statement: The Bayesian viewpoint is very likely the most relevant probabilistic framework for science.

\end{multicols}

\begingroup
\parindent 0pt
\parskip 1pt
\def\enotesize{\small}
\theendnotes
\endgroup

\end{document}